\documentclass{multi3}
\usepackage{graphicx}
\usepackage{amsmath} 

\font\cal=cmsl10  
\font\smallrm=cmr8 
\newcommand\Beta{{\rm Beta}}
\newcommand\Dir{{\rm Dir}}
\newcommand\E{{\rm E}} 
\newcommand\N{{\rm N}} 
\newcommand\RR{\mathord{I\kern-.3em R}}
\newcommand\Var{{\rm Var}} 
\newcommand\tr{{\rm t}} 
\renewcommand\d{{\rm d}} 
\newcommand\data{{\rm data}}
 
\newcommand\midd{\,|\,}  
\newcommand\arr{\rightarrow} 
\newcommand\hatt{\widehat} 
\newcommand\tilda{\widetilde}  
 
\newcommand\sumin{\sum_{i=1}^n}

\newcommand\eps{\varepsilon} 
\newcommand\half{\hbox{$1\over2$}}

\newcommand\LL{L\'evy} 
\newcommand\Bernstein{{Bernshte\u\i n}} 

\begin{document}
\chapter{Topics in Nonparametric Bayesian Statistics}
\author{Nils Lid Hjort}
\address{University of Oslo}

\section{Introduction and summary} 

The intersection set of Bayesian and nonparametric
statistics was almost empty until about 1973, 
but now seems to be growing at a healthy rate.
This chapter gives an overview of various theoretical and applied 
research themes inside this field, partly complementing 
and extending recent reviews of Dey, M\"uller and Sinha (1998) 
and Walker, Damien, Laud and Smith (1999). 
The intention is not to be complete or exhaustive, 
but rather to touch on research areas of interest, partly by example. 

\subsection{What is it, and why?} 

In this chapter we do not use a very precise definition 
of what constitute `nonparametric Bayesian methods', and might 
err on the liberal side. Specifically, examples are included 
of statistical modelling and inference situations placing 
a distribution over large sets of probability distributions
is one of the ingredients. Some of these situations do not 
have to be intrinsically Bayesian per se. 

One of the goals of nonparametric Bayesian statistics is 
to ease up on traditional `hard' model assumptions, without 
essential loss of inference power. Pure finite-parametric models 
can never be fully correct, whereas nonparametric Bayes 
constructions may succeed in having most conceivable 
true data generating mechanisms inside its prior scope,
i.e.~its support. If the setup is satisfactory, and the data 
quality reasonable, one often finds that the data themselves 
help dictate to what extent solutions are close or not close
to what they would have been under simpler assumptions. 
Such findings, along with easily available software tools 
that most statisticians can learn to use, make up good selling points.
These `harder' model assumptions of traditional statistics, 
both frequentist and Bayesian, might include both the error and 
the signal structure of models. Thus one might soften up the 
linear normal regression textbook methods by using a nearly linear 
mean function, nearly Gau{\ss}ian errors, nearly constant variance
level across covariates, and if relevant some dependence 
structure. This also serves to illustrate that there by necessity 
is a broad range of possible nonparametric constructions, 
where one should not anticipate clear winners.

The essence of the `nonparametric' word is that what is being modelled
is not seen as well enough described by a fixed (and perhaps low) 
number of parameters. Otherwise the term in current usage is 
not very strictly interpreted. It might allude to broad flexibility 
(many shapes of the underlying curve or surface being possible 
under the model), which can be achieved in several ways. 
Some constructions use a growing number of parameters, 
or perhaps a growing range of candidate models to choose from
or average over, e.g.~involving mixtures. Here the operating
key word is flexibility, there being no clear division between 
parametric and nonparametric; see also Green and Richardson (2001). 
The nonparametric term might also allude to certain mathematical 
aspects of performance as the data volume grows, like 
consistency or optimality of precision.

Statistics has witnessed the three first decades of nonparametric
Bayesian life, which arguably has passed through its infancy 
and early youth. The first period was primarily a mathematical 
or probabilistical one, by necessity concerned with setting up 
the right probability structures on the proper spaces and
deducing, when possible, relevant aspects of the posterior
distributions. The second period has been more explorative, 
making different approaches more flexible and amenable to practical 
analysis in a growing list of applications. This has also, 
through serendipitous timing, been aided by broadly enhanced
computing abilities and methodology, including software and bigger
toolboxes for stochastic simulation, in particular 
Markov chain Monte Carlo methods (see e.g.~chapters by Roberts 
and Green, this volume, with discussions), along with more frequent 
use of numerical analysis software. Of course both `periods' 
are in a sense never-ending stories.

In spite of broad impressive developments many nonparametric Bayes
setups will continue to pose challenges of construction, deduction,
interpretation and computation. Given these complexities, related 
also to probability calculus over infinite-dimensional spaces, 
it is not surprising that a fair portion of published work in this area 
has been in the `can do' spirit. This is also true for many 
applications to real data. For the envisaged upcoming third period 
in the life of nonparametric Bayes one might predict further broadening 
and maturing of the field, leading with experience to more finesse 
and possibly a higher degree of scientific relevance in old
and new segments of substantial statistical application. 
At the same time more theory and a broader range of models 
will be developed. It is also likely that more hybrid constructions
will evolve, perhaps mixing together not only parametric and 
nonparametric ingredients for given problems, but also by perhaps 
pragmatic frequentist-inspired solutions to aid constructions 
that at the outset are meant as pure Bayesian. There will be
challenges of combining different data sources of different 
quality and complexity, where nonparametric Bayes might play
a role, but along with other elements. 
Efron (2002) predicts a wave of empirical Bayes statistics 
for the 21st century, for example in connection with problems 
of microarrays and data mining for big databases.
This wave should also encompass empirical nonparametric Bayes methods. 

\subsection{On the present chapter} 

In Section 2 the Dirichlet process is proved to be a distributional 
limit of certain simpler processes with symmetrically distributed 
probability weights. This suggests suitable generalisations
of the Dirichlet for use as priors in Bayesian inference,
and also proves useful in connection with transform identities
for distributions of random means, as shown in Section 3. 
Identities are obtained there which partly generalise earlier results
of Cifarelli and Regazzini (1990) and Diaconis and Kemperman (1996). 
Section 2 also provides another generalisation of the Dirichlet,
starting with the infinite series representation due to 
Sethuraman and Tiwari (1982). 
Section 4 deals with quantile inference, first based on
the Dirichlet process and then using a more general nonparametric 
prior quantile process, which is constructed in a pyramidal fashion.
These quantile trees aim at being for the quantiles what the 
P\'olya trees are for the cumulative distribution functions.  
A natural quantile function estimator is seen to lead to
an attractive Bayesian density estimator, which does not require
any smoothing parameters. Some interpretational
and consistency issues are then discussed for 
Bayesian density estimation in general. 

Section 6 shows how elements of nonparametric Bayesian modelling 
may be used for a different purpose than merely analysing data, 
namely to derive functional forms of statistical functions
in regression contexts. It is shown there how a broad class 
of \LL{} type cumulative damage processes, viewed as frailty
processes for individuals, influence their survival distributions
in a way leading to the multiplicative hazard regression structure. 
Then in Section 7 we briefly discuss the use of Beta processes
to a linear hazard regression model, before going on in Section 8 
to a broad class of Bayesian extensions of the by now traditional
way of carrying out nonparametric regression, namely that of 
local polynomial modelling. In Section 9 a use is found for 
smoothed Dirichlet processes as a modelling tool for random shapes,
and in Section 10 nonparametric envelopes around parametric models
are studied. Finally Section 11 offers some concluding remarks. 

\section{The Dirichlet process prior and some extensions} 

The Dirichlet process prior was introduced in Ferguson (1973, 1974) 
and remains a cornerstone in Bayesian nonparametric statistics. 
It is also a favourite special case of various generalisations
that have been worked with, including neutral to the right 
and tailfree processes (Doksum, 1974, Ferguson, 1974), 
P\'olya trees (Kraft, 1964, Ferguson, 1974, Lavine, 1992), 
Beta processes (Hjort, 1990) and mixtures of Dirichlets 
(Lo, 1984, Escobar and West, 1995). Below we establish some notation,
review the Dirichlet, and briefly discuss two useful extensions.

\subsection{The Dirichlet process}
 
To define the Dirichlet process on a sample space $\Omega$,
let $P_0$ be a probability measure thereon, interpreted as 
the prior guess distribution for data, and let $b$ be positive. 
Then $P$ is a Dirichlet process with parameters $(b,P_0)$, 
for which we write $P\sim\Dir(b,P_0)$, 
if for each partition $A_1,\ldots,A_k$,
$$\bigl(P(A_1),\ldots,P(A_k)\bigr)
  \sim\Dir\bigl(bP_0(A_1),\ldots,bP_0(A_k)\bigr). \eqno(2.1)$$
We may refer to $bP_0$ as the total measure. 
In particular, for each set $A$, $P(A)\sim\Beta\{bP_0(A),b(1-P_0(A))\}$
with mean $P_0(A)$ and variance $P_0(A)(1-P_0(A))/(1+b)$. 
Perhaps the most attractive property of the Dirichlet prior
is the ease with which it is being updated with incoming data;
if $x_1,\ldots,x_n$ are observations having arisen as 
an independent $n$-sample from the random $P$, then
$P$ given these is another Dirichlet, with total measure 
$bP_0+n\hatt P_n$. Here $\hatt P_n$ is the empirical distribution 
giving mass $1/n$ to each data point. 

One often refers to the limiting case $b\arr0$,
where $P$ is concentrated at the data points with 
probabilities given by a flat Dirichlet $(1,\ldots,1)$,  
as corresponding to using a non-informative prior for $P$. 
There are many cases where Bayesian inference using
this posterior gives natural parallels to perhaps canonical
frequentist methods; cases in point include the empirical
distribution $\hatt P_n$ as limiting Bayes estimate, 
and the Bayesian bootstrap developed by Rubin (1981) and others. 
See also Sections 4.1--4.2 below. The notion of non-informativity
is debatable here, however, as the behaviour of 
the prior process $P$ is peculiar when $b$ is small.
In the limit, it is concentrated at a single value, chosen from $P_0$. 

\subsection{The Dirichlet as a limit}
 
Hjort and Ongaro (2002) give a new constructive definition of 
the Dirichlet process as a limit of simpler processes. Let 
$$P_m=\sum_{j=1}^m\beta_j\delta(\xi_j)
  \quad {\rm where\ }\beta=(\beta_1,\ldots,\beta_m)
  \sim\Dir(b/m,\ldots,b/m), \eqno(2.2)$$
where $\xi_1,\xi_2,\ldots$ are independent from $P_0$ 
and independent of $\beta$. Here $\delta(\xi)$ denotes
unit point mass at position $\xi$. For a set $A$, 
and conditionally on the $\xi_j$s, $P_m(A)$ is a Beta
with parameters $\{b\hatt P_m(A),b(1-\hatt P_m(A))\}$, 
where $\hatt P_m$ is the empirical distribution of $\xi_1,\ldots,\xi_m$. 
Hence $P_m(A)$ is distributed as a binomial mixture over 
such Beta distributions. Since $\hatt P_m(A)$ goes to $P_0(A)$
as $m\arr\infty$ and the Beta is continuous in its parameters, 
the limit distribution of $P_m(A)$ is the Beta distribution of $P(A)$
when $P\sim\Dir(b,P_0)$. An extension of this argument shows
that all finite-dimensional distributions of $P_m$ go to those given in (2.1). 
An interesting connection is that $\Pr\{P_m(A)\in C\}$,
for any set $C$, can be seen to be the \Bernstein{} 
polynomial approximation (see e.g.~Billingsley, 1995)
to the function $h(p)=\Pr\{\Beta(bp,b(1-p))\in C\}$ 
at the point $p=P_0(A)$. 
Proving convergence of $P_m$ can also be done via
results about the so-called Poisson--Dirichlet distribution,
see Kingman (1975), and which also has connections to
representation (2.3) below. 

The limit construction (2.2) is different from but shares some of the 
ingredients of the infinite series representation considered below.
Among its advantages is the simplicity of the symmetric 
Dirichlet for the weights. In addition to being useful for deriving
facts about the Dirichlet, as indicated in the next section, it 
also invites suitable bona fide generalisations of the Dirichlet
process prior. A simple construction of interest is to let 
$P=\sum_{j=1}^M\beta_j\delta(\xi_j)$, where $M$ has any distribution
on the integers with $\Pr\{M>m\}$ positive for all $m$. 
One may develop methods for Bayesian inference using this 
nonparametric prior. The Dirichlet is the limiting case where 
$M$ tends to infinity.

\subsection{An extension via an infinite sum representation} 

Consider independent $B_1,B_2,\ldots$ drawn from the same distribution
$H$ on $(0,1)$. These generate random probabilities 
$\gamma_1=B_1$, $\gamma_2=\bar B_1B_2$, $\gamma_3=\bar B_1\bar B_2B_3$
and so on, where $\bar B_j=1-B_j$; these sum a.s.~to 1 since 
$1-\sum_{j=1}^n\gamma_j=\bar B_1\cdots\bar B_n$. Accordingly we 
may define a random probability measure by 
$$P=\sum_{j=1}^\infty\gamma_j\delta(\xi_j)
  \quad {\rm with\ }\xi_1,\xi_2,\ldots {\rm i.i.d.}\sim P_0. \eqno(2.3)$$
Sethuraman and Tiwari (1982) showed that the Dirichlet process $(b,P_0)$
can be represented in this form, for the particular choice of 
a $\Beta(1,b)$ distribution for the $B_j$s; see also Sethuraman (1994).
Ishwaran and Zarepour (2000) and Hjort (2000) have 
independently studied the extension to a general distribution $H$ 
for these. A fruitful family of priors emerges by letting 
$H=\Beta(a,b)$, creating a generalised Dirichlet process 
with parameters $(a,b,P_0)$. Ishwaran and Zarepour (2000)
develop computational algorithms using simulation, while
more explicit theoretical results about estimators 
and performance are reached in Hjort (2000). 
Note that the Dirichlet corresponds to $a=1$, an inner point
in the parameter space of its generalisation. This extension
allows more modelling flexibility regarding the skewness,
kurtosis and so on of random means. 
Explicit formulae are available for posterior means 
and variances of random mean parameters. One finding of 
general importance is that the speed with which the data
wash out aspects of the prior is of the order $O(n^{-a})$,
which can be slower or faster than the ordinary rate 
$n^{-1}$ found for nearly all parametric problems 
as well as for the Dirichlet prior. 

\section{Random Dirichlet means} 

For $P$ a Dirichlet process $(b,P_0)$ on a sample space $\Omega$,
consider a random mean 
$\theta=\E_Pg(X)=\int g\,\d P$. 
There are many uses of such constructions besides 
the most immediate one where it is a focus parameter for Bayesian inference. 
Recently much attention has been given to the study of 
the distribution of such a $\theta$; 
a partial list is Diaconis and Kemperman (1996), 
Regazzini, Guglielmi and di Nunno (2000), Hjort and Ongaro (2002), 
Guglielmi and Tweedie (2000) and Tsilevich, Vershik and Yor (2000). 

\subsection{Transform identities} 

The task is to derive aspects of the distribution of $\theta$
using information about $Y=g(\xi)$, where $\xi\sim P_0$. 
Assume for simplicity of presentation that $g$ is nonnegative. 
Cifarelli and Regazzini (1990, 1994) were the first to exhibit
an identity linking the so-called Hilbert transform of $\theta$
to a transform of $Y$. This connection may be written 
$$\E\exp\Bigl\{-b\log\Bigl(1+u\int g\,\d P\Bigr)\Bigr\}
   =\exp\Bigl[-b\int\log\{1+ug(\xi)\}\,\d P_0(\xi)\Bigr]. \eqno(3.1)$$ 
Cifarelli and Regazzini gave a rather long proof of (3.1) 
and some of its variations, and used integration 
in the complex plane to indicate how the transform may be inverted 
to find the distribution of $\theta$ numerically. 

A quite straightforward derivation of (3.1), and without unnecessary 
side conditions, are among the consequences of construction (2.2) 
discussed in Hjort and Ongaro (2002). 
One may write $\beta_j=G_j/S_m$ in terms of independent 
Gamma $(b/m,1)$ variables $G_1,\ldots,G_m$ and their sum $S_m$. 
Write therefore $\theta_m=\theta(P_m)=R_m/S_m$, with $R_m=\sum_{j=1}^mG_jY_j$ 
being a random mixture of many small Gammas; here, $Y_j=g(\xi_j)$. 
First, exploit independence between $\theta_m$ and $S_m$ to derive 
$$\E\exp(-uR_m)=\E[\E\exp(-u\theta_mS_m)\midd S_m]
   =\E\exp\{-b\log(1+u\theta_m)\}, $$
using the fact that $S_m$ has Laplace transform $(1+u)^{-b}$. 
Next, use the Laplace transform $(1+u)^{-b/m}$ for the $G_j$s to obtain
$$\E[\exp(-uR_m)\midd\xi_1,\ldots,\xi_m]
  =\prod_{j=1}^m(1+uY_j)^{-b/m}
  =\exp\Bigl\{-b{1\over m}\sum_{j=1}^m\log(1+uY_j)\Bigr\}. $$
Taking the limit, and supplying some extra arguments, 
one proves (3.1); both sides are equal to the Laplace transform 
of the variable $R$ to which $R_m$ converges in distribution. 
Hjort and Ongaro also give a multivariate version of this,
in the form of a formula for $\E\exp\{-b\log(1+\sum_{j=1}^ku_j\theta_j)\}$,
where $\theta_j=\int g_j\,\d P$. Such results were 
explicitly mentioned as missing in the literature by Diaconis 
and Kemperman (1996). See also Kerov and Tsilevich (1998). 

Let $G$ be a Gamma process on the sample space with parameter $bP_0$;
it has independent contributions for disjoint sets and 
$G(A)$ is Gamma $(bP_0(A),1)$ for each $A$. 
The arguments used above are connected to 
the representation of a Dirichlet process as a normalised Gamma process, 
viz.~$P(\cdot)=G(\cdot)/G(\Omega)$, where one in addition may demonstrate 
that $P(\cdot)$ is independent of $G(\Omega)$. Given the simplicity
of these arguments it is perhaps not surprising that several 
authors recently and independently have come up with somewhat different 
but related proofs of (3.1) and its relatives; 
in addition to Hjort and Ongaro (2002),
see Regazzini, Guglielmi and di Nunno (2000) and 
Tsilevich, Vershik and Yor (2000). 
One may in fact trace the roots of identity (3.1) back to Markov (1896). 

\subsection{Stochastic equations and the full moment sequence} 

The mean of $\int g\,\d P$ is $\theta_0=\int g\,\d P_0$,
and Ferguson (1973) gave a formula for the variance. 
Among the uses of (2.2) and (3.1) is the possibility of deriving 
formulae for further moments; see Hjort and Ongaro (2002)
for a list of the first ten centralised moments $\E(\theta-\theta_0)^p$. 
One may also derive a stochastic equation for the distribution of $\theta$,
as follows. Use representation (2.3) to write 
$\theta=\sum_{j=1}^\infty\gamma_jY_j$
in the form $B_1Y_1+\bar B_1(B_2Y_2+\bar B_2B_3Y_3+\cdots)$,
from which it is apparent that 
$$\theta=_dBY+\bar B\theta. \eqno(3.2)$$
On the right hand side, $B\sim H$, $Y=g(\xi)$ with $\xi\sim P_0$
and $\theta$ are independent, and `$=_d$' indicates 
equality in distribution. This stochastic equation 
determines the distribution of $\theta$ uniquely. 
Only rarely can this distribution be exhibited in closed form,
however. The equation at least gives a simple recursive method 
of finding all moments, via 
$$\E(\theta-\theta_0)^p
   =\sum_{j=0}^p{p\choose j}\E_0(Y-\theta_0)^{p-j}
    \,\E B^{p-j}\bar B^j\,\E(\theta-\theta_0)^j
   \quad {\rm for\ }p\ge 2. \eqno(3.3)$$
Here `$\E_0$' indicates expected value when $Y$ has 
its null distribution $Q_0=P_0g^{-1}$. 
Note that this gives a recipe for finding all moments not only 
for the Dirichlet case, where $B\sim\Beta(1,b)$, but also
for the generalised process of (2.3) where $B$ has an arbitrary 
distribution on $(0,1)$. 

It is not difficult to use equation (3.2) to construct
a Markov chain with the distribution of $\theta$ as
its equilibrium. Such simulation output can be further
repaired to give improved accuracy via knowledge of 
the exact moments, as demonstrated in Hjort and Ongaro (2002). 

\section{Quantile pyramid processes} 

Let $Q(y)=F^{-1}(y)$ be the quantile function for a 
distribution on the real line, and suppose data are observed 
from this distribution. One may attempt to carry out quantile inference
via a given nonparametric prior for $F$, and this is done
below for the Dirichlet case. It is also worthwhile to place priors 
directly on the set of quantile functions, leading to direct Bayes 
estimators of $Q$ and related functions. 

\subsection{Quantile inference with the Dirichlet process} 

Let $F$ be the cumulative function of a $\Dir(b,F_0)$ process, 
where $F_0$ is a suitable prior guess distribution with density $f_0$, 
and define more formally 
$$Q(y)=F^{-1}(y)=\inf\{t\colon F(t)\ge y\} 
   \quad {\rm for\ }y\in(0,1). \eqno(4.1)$$
For this left-continuous inverse of the right-continuous $F$ it holds
that $Q(y)\le x$ if and only if $y\le F(x)$. It follows that 
the distribution of $Q(y)$, prior to data, is given by
$$\Pr\{Q(y)\le x\}=1-G(y;bF_0(x),b\bar F_0(x))
   =G(1-y;b\bar F_0(x),bF_0(x)), \eqno(4.2)$$
where $G(y;a,c)$ is the distribution function for a $\Beta(a,c)$
and $\bar F_0=1-F_0$. Note that (4.2) may be written $J_b(F_0(x))$, 
where $J_b(x)=G(1-y;b(1-x),bx)$ is the distribution of a random
$y$-quantile for the special case of $F_0$ being uniform on $(0,1)$. 
It also follows that $Q(y)$ has a density of the form 
$j_b(F_0(x))f_0(x)$, where $j_b=J_b'$ is the density under the 
uniform prior measure. 

The above immediately gives results for the posterior distributions 
of quantiles given a set of data $x_1,\ldots,x_n$, in view of 
the updating mechanism for the Dirichlet. Assume for simplicity
of presentation that data points are distinct and rank them to 
$x_{(1)}<\cdots<x_{(n)}$, and add on $x_{(0)}=-\infty$ 
and $x_{(n+1)}=\infty$. Then, for $x_{(i)}\le x<x_{(i+1)}$, 
$$H_{n,b}(x)=\Pr\{Q(y)\le x\midd\data\}
   =G(1-y;b\bar F_0(x)+n-i,bF_0(x)+i). $$
It has a suitable density inside data windows $(x_{(i)},x_{(i+1)})$
and point masses 
\begin{equation} 
\begin{split} 
\Delta H_{n,b}(x_{(i)})
&= G(1-y;b\bar F_0(x_{(i)})+n-i,bF_0(x_{(i)})+i) \cr
&  \qquad\qquad 
  -\,G(1-y;b\bar F_0(x_{(i)})+n-i+1,bF_0(x_{(i)})+i-1) \cr
&= {\rm const.} 
  \,y^{bF_0(x_{(i)})+i-1}(1-y)^{b\bar F_0(x_{(i)})+n-i} 
\nonumber 
\end{split}
\end{equation} 
at point $x_{(i)}$. In the case $b\arr0$ there is no posterior probability
left between data points; the posterior of $Q(y)$ concentrates on
the data points with probabilities 
$$\Delta H_{n,0}(x_{(i)})=p_{n,y}(x_{(i)})
   ={n-1\choose i-1}y^{i-1}(1-y)^{n-i}
   \quad {\rm for\ }i=1,\ldots,n. \eqno(4.3)$$
The Bayesian quantile estimator function is 
$\hatt Q_b(y)=\E\{Q(y)\midd\data\}$. The non-informative limit 
is of particular interest here: 
$$\hatt Q_0(y)=\sum_{i=1}^n{n-1\choose i-1}y^{i-1}(1-y)^{n-1}x_{(i)}
   \quad {\rm for\ }y\in(0,1). \eqno(4.4). $$
This estimator can be seen as a \Bernstein{} polynomial approximation
to a version of the empirical quantile estimator. It has been 
worked with earlier by Cheng (1995) and others, but is here 
given additional interpretational weight as the non-informative limit 
of a natural nonparametric Bayesian estimator. 
Issues related to this are further discussed in 
forthcoming work with S.~Petrone. 
They also exhibit the full posterior process $Q(\cdot)$,
as opposed to concentrating on a single $y$ at a time. 

\subsection{An automatic nonparametric density estimator} 

Note that $\hatt Q_b(y)$ and $\hatt Q_0(y)$ are smooth 
estimates of $F^{-1}(y)$, whereas the corresponding Bayes estimators 
$\hatt F_b(t)$ and $\hatt F_0(t)$ for $F$ under quadratic loss 
have jumps at the data points. 
One may take the derivative to get an estimate of 
the quantile density function $q(y)=1/f(F^{-1}(y))$. 
This may be inverted to find a density estimate 
$\hatt f_b(x)=1/\hatt q_b(\hatt F_b(x))$. 
It requires numerically solving 
$\hatt Q_b(\hatt F_b(x))=x$ for $\hatt F_b(x)$, for each $x$. 
A particularly attractive automatic nonparametric density estimator
emerges when $b\arr0$. 
The resulting $\hatt f_0(x)$ does not require a separate
smoothing parameter. It is zero outside the data range 
$[x_{(1)},x_{(n)}]$, with 
$$\hatt f_0(x_{(1)})={1\over (n-1)(x_{(2)}-x_{(1)})} 
   \quad {\rm and} \quad 
  \hatt f_0(x_{(n)})={1\over (n-1)(x_{(n)}-x_{(n-1)})}, $$
and is positive and smooth inside. For large $n$ it is approximately
equal to a kernel type density estimator with a Gau\ss ian kernel
and variable bandwidth proportional to $n^{-1/2}$. 

\subsection{Quantile pyramids} 

\def\quart{\hbox{$1\over 4$}}
\def\threequart{\hbox{$3\over 4$}}
The following is an attempt to construct a prior process 
$Q(\cdot)$ directly on the set of quantile functions. 
Let us for convenience work on distributions on the unit interval $[0,1]$, 
so that $Q(0)=0$ and $Q(1)=1$. The starting point is a class 
of distributions that can be specified on arbitrary bounded intervals. 
Let $p_{m,[a,b]}$ denote a density concentrated at 
sub-interval $[a,b]$, to be employed at level $m$ in a growing 
pyramid, or tree. A simple possibility is to fix a density $h$ 
on the broadest interval in question and then scale it to 
$h(x)/\int_a^b h(x)\,\d x$ on the required sub-interval. 
To describe the intended prior quantile process, 
first draw the median $Q(\half)$ from distribution $p_{1,[0,1]}$, 
say. Then draw the quartiles independently, say 
$Q(\quart)\sim p_{2,[0,Q(1/2)]}$ and $Q(\threequart)\sim p_{2,[Q(1/2),1]}$. 
At stage three one draws the four remaining octiles 
$Q({1\over 8}),Q({3\over 8}),Q({5\over 8}),Q({7\over 8})$ 
independently from the appropriate $p_{3,\cdot}$ distributions
on the appropriate intervals, and so on. At stage $m$ 
new quantiles $Q(j/2^m)$ are generated for $j=1,3,\ldots,2^m-1$, 
conditional on the values already generated
at levels 1 to $m-1$ above, and $Q(j/2^m)$ depends only upon 
its two parents $Q((j\pm1)/2^m)$. 
In this way a `quantile pyramid' or `quantile tree' is grown. 
The construction resembles that of P\'olya trees, see 
Ferguson (1974), Lavine (1992, 1994) and 
Walker, Damien, Laud and Smith (1999), 
but is different in spirit and operation. 
With P\'olya trees the partitions are fixed (as dyadic intervals) 
but the probabilities are random (as Beta variables); 
this specifies a random distribution function $F$. 
Here we fix probabilities instead (in the natural dyadic fashion)
and use random partitions; the result is the quantile function $Q$. 

The quantile pyramid may be stopped at some level $m$,
after which linear interpolation defines the remaining
parts of the distribution. It may also be allowed to go on
indefinitely to define a full stochastic process $Q$ 
on $(0,1)$ not determined from a finite number of parameters,
thus constituting a genuine nonparametric prior quantile process. 
Existence of the process follows by tightness of the sequence
of finite approximations.
The quantile-Dirichlet process touched on above can be shown 
to be a special case.

\subsection{Posterior quantile pyramids} 

Assume data $x_1,\ldots,x_n$ have been observed.
The challenge is to determine the behaviour 
of the posterior quantile process. 
One point of view is that a $Q$ process determines an $F$ 
for which general principles for finding the posterior of $F$ apply; 
hence $Q=F^{-1}$ may be analysed too. This is often complicated, however,
and the following two alternatives appear quite fruitful. 

Assume for illustration of the first idea that 
there is a simultaneous density for the 15 sedecimiles 
$q_j=Q(j/16)$ of the form given above, and that $Q$ otherwise 
is defined by linear interpolation. This means that 
its $F$ is also linear over the 16 intervals defined by the 
15 quantiles, that is, the distribution has a constant density
$F'(x)={1\over 16}/(q_j-q_{j-1})$ for $x\in (q_{j-1},q_j)$, 
for each of the 16 sub-intervals. This gives a likelihood for the data
proportional to 
$$L_{n,1}(q)=\prod_{j=1}^{16}\Bigl({1\over q_j-q_{j-1}}\Bigr)^{N_j(q)}
  \quad {\rm for\ }q_1<\cdots<q_{15}, $$
where $N_j(q)=nF_n(q_{j-1},q_j)$ is the number of data points 
falling in quantile-defined $x$-interval number $j$. 
This makes it possible to write down the posterior density of 
$(q_1,\ldots,q_{15})$. Algorithms of Metropolis--Hastings type 
may be put up to simulate from this distribution;
see Hjort and Walker (2002). 

A second route is that offered by what may be termed the 
substitute likelihood. In the setting above, assume that a 
pyramid-type prior is given for the 15 quantiles $q_1,\ldots,q_{15}$,
but we avoid any further specification of $Q$. The substitute
likelihood for data, say $L_{n,2}(q)$, is the multinomial probability 
$${n\choose N_1(q),\ldots,N_{16}(q)}
        \Bigl({1\over 16}\Bigr)^{N_1(q)}\cdots
        \Bigl({1\over 16}\Bigr)^{N_{16}(q)}
      ={n!\over N_1(q)!\cdots N_{16}(q)!}\Bigl({1\over 16}\Bigr)^n. $$

With the substitute likelihood and a pyramid type prior for the 
quantiles there is a convenient way of expressing the 
(substitute-based) posterior density, as shown in Hjort and Walker (2002).
The point is to rearrange the multinomial terms to match 
the tree-structure of the prior. Here it means that the 15 
quantiles follow the same pyramidal dependence structure 
given data as they did in the prior. 
This partial conjugacy type result has the practical advantage that 
one may deal with one quantile at the time, following the tree.
First one simulates a median from $p(q_8\midd\data)$,
then the two quartiles from respectively $p(q_4\midd\data,q_8)$
and $p(q_{12}\midd\data,q_8)$, and so on. These individual 
simulation steps could use a Metropolis--Hastings type algorithm. 
Repeating the full process many times over gives in the 
end posterior distributions for quantiles of interest.
This rearrangement can actually also be carried out for 
the first linear interpolation likelihood, and indeed 
$L_{n,1}$ and $L_{n,2}$ can be shown to behave similarly for large $n$. 

\section{Bayesian density estimation and consistency issues} 

Nonparametric Bayesian density estimation means placing a 
prior distribution on the set of densities and then 
analysing aspects of the posterior distribution;
an overview with many as yet not fully explored 
approaches is in Hjort (1996). 
A technical point worth mentioning is that one may compute 
the posterior mean and variance functions via simulations
from the prior alone, that is, without having to assess
full aspects of the posterior distribution of the density. 

Assume a prior $\pi$ is constructed for an unknown
continuous density $f$. If the data really follow
a density $f_0$, will the posterior distribution
$\pi(f\midd{\rm data})$ concentrate around $f_0$,
as the sample size increases? This topic is currently a hectic one,
and various authors have reached different, highly technical 
and perhaps rather harsh sets of conditions sufficient to ensure consistency; 
see Wasserman (1998), Barron, Schervish and Wasserman (1999), 
Ghosal, Ghosh and Ramamoorthi (1999), Shen and Wasserman (2000) 
and Ghosal and van der Vaart (2000). 

An important result was established as early as Schwartz (1965).
She showed that under a condition which we will denote (A), 
and which is that $\pi$ puts positive mass on all 
Kullback--Leibler neighbourhoods $\{f\colon\int f_0\log(f_0/f)<\eps\}$ 
around $f_0$, then the posterior is at least weakly consistent. 
This means that for almost all sequences under $f_0$, 
$\pi(U\midd\data)\arr1$ for all weak neighbourhoods 
$U=\{f\colon w(F_0,F)<\eps\}$ around $f_0$; here $w$ is any
metric on the cumulatives $F_0$, $F$ equivalent to convergence
in distribution. Strong Hellinger consistency demands more, namely that 
$\pi(U\midd\data)$ should a.s.~go to 1 also for the potentially 
much more complicated neighbourhoods $U=\{f\colon H(f_0,f)<\eps\}$,
where $H(f_0,f)^2=\int(f^{1/2}-f_0^{1/2})^2\,\d x=2-2\int(f_0f)^{1/2}\,\d x$. 
Consistency is a statement concerning the pair $(\pi,f_0)$;
one typically wishes conditions under which a prior $\pi$ 
gives consistency for large sets of $f_0$. 
Conditions ensuring strong consistency given in the many 
recent papers on the subject typically take the form `(A) and (B)',
where (A) is the minimum requirement given above and where 
(B) varies in content, sharpness and context from one article to another. 

Here we make two points. The first is in the technical tradition 
and holds that versions of condition (B) given in several recent 
articles have been too strong. Walker and Hjort (2001) work with 
sequences of suitably modified posteriors and show that these 
are truly strongly consistent under condition (A) alone. 
The modification in question may be seen as having arisen
either from a modification of the prior or from a robustification
of the likelihood function. A corollary gives easy and weak
conditions for the Bayes estimator (posterior mean) to be 
Hellinger consistent. This provides a circumventive way 
of establishing strong consistency for the sequence of real posteriors
for many classes of prior distributions. 

To give one illustration, consider a P\'olya tree
prior employing $\Beta(a_k,a_k)$ variables at level $k$; 
an old result of Kraft (1964) guarantees that the randomly chosen $F$ 
a.s.~has a density $f$ as long as $\sum_{k=1}^\infty 1/a_k^{1/2}$
converges. As long as this holds 
and the Kullback--Leibler divergence between $f_0$ 
and the prior predictive is finite, condition (A) holds.  
The (B) condition used by Barron, Schervish and Wasserman (1999)
leads to the very strict criterion $a_k=8^k$ (or even faster),
which means P\'olya trees where the Beta components
become almost pre-determined even for low $k$,
i.e.~trees with leaves that hardly move after three or four levels. 
With the Walker and Hjort (2001) strategy, however, 
the condition $\sum_{k=1}^\infty 1/a_k^{1/2}<\infty$ 
alone is sufficient to secure Hellinger consistency of 
the predictive distribution; for example, it suffices to have 
$a_k$ of the type $ck^{2+\eps}$ for some positive $\eps$. 

The second point worth raising here is that for most statistical 
and decision related applications one would be quite content
with weak consistency, which is secured under the basic 
nonparametric prior condition (A) alone. One may argue that 
with weak consistency one learns the true cumulative for large $n$,
and this suffices to learn also the derivative, even 
in the few and rather special situations where strong
Hellinger consistency fails. Walker (2000) discusses similar points.

\section{\LL{} frailty processes and proportional hazards}

The assumption of proportional hazards functions plays a major role in
survival and event history analysis. Judged by the extremely wide
application of methods based on proportional hazards, especially 
in terms of Cox models, it seems clear that one ought to understand 
better what this assumption really means. For instance, 
when assuming proportional hazards this is a statement
about averages: on the `average' the hazard in a group is, say, 
twice the hazard in another group. However, each group will contain 
a wide variation in individual risk, and one may ask what proportional 
hazards means for this variation. This, of course, is a frailty point 
of view. Although frailty considerations often lead to prediction of 
decreasing hazard ratios, this is not always so. 

Aalen and Hjort (2002) present classes of frailty constructions
which necessarily lead to proportional hazards. The approach 
taken in that article is not Bayesian per se, but the classes 
worked with rely on probability measures being constructed on large sets
and have interpretations in Bayesian terms. One construction,
complementing that of Aalen and Hjort, is as follows. 
Individuals are pictured as being continuously exposed 
to an unobserved cumulative damage type process, of the form 
$$Z(t)=\sum_{j\le M(t)} \theta G_j 
  \quad {\rm for\ }t\ge 0. \eqno(6.1)$$ 
Here $G_1,G_2,\ldots$ are taken to be i.i.d.~nonnegative variables, 
interpreted as adding over time to the hazard level of the individual, 
while $M(\cdot)$ is a Poisson process with cumulative rate 
$\Lambda(t)=\int_0^t\lambda(s)\,\d s$, that is, its increments 
are independent and Poisson $\lambda(s)\,\d s$. 
The $\theta$ is an additional parameter acting multiplicatively 
on the $G_j$s. There is a certain over-parameterisation in that $\theta$ 
may be subsumed into the $G_j$s in (6.1), but it is convenient
for later modelling purposes to keep it present. 
From a modelling perspective one may work from different
sets of assumptions about the $G_j$ distribution, 
or the Poisson intensity $\lambda(t)$, or the $\theta$ factor,
depending in suitable ways on covariate information. 

\def\hist{{\cal H}}
The specific connection to the person's survival prospects is to 
model $S(t\midd\hist_t)\allowbreak=\Pr\{T\ge t\midd\hist_t\}$, the survival
distribution given the full history of what has happened to 
the person up to time $t-$, as 
$$S(t\midd\hist_t)=\exp\{-Z(t)\}=\prod_{j\le M(t)}\exp(-\theta G_j)
   =\prod_{j\le M(t)}(1-R_j)^\theta. \eqno(6.2)$$
Here $R_j=1-\exp(-G_j)$. Letting $L_0(u)=\E\exp(-uG_j)$ 
be the Laplace transform of the $G_j$s, it follows that 
the unconditional survival function must take the form
$$S(t)=\E\exp\{-Z(t)\}=\E L_0(\theta)^{M(t)}
   =\exp[-\Lambda(t)\{1-L_0(\theta)\}]. \eqno(6.3)$$ 
Note that even though the survival function is discontinuous
given the jumps of the unobservable damage process, 
it becomes continuous marginally, with cumulative hazard rate
$H(t)=\Lambda(t)\{1-L_0(\theta)\}$ and hazard rate function
$$h(s)=\lambda(s)\{1-L_0(\theta)\}
   =\lambda(s)\{1-\E(1-R)^\theta\}. \eqno(6.4)$$
One may now add aspects of observable covariate information
on to the framework above. 
Consider individuals $i=1,\ldots,n$ with covariate vectors $x_1,\ldots,x_n$.  
Equation (6.2) translated to individual $i$ holds that 
$S(t\midd x_i,Z_i)=\exp\{-Z_i(t)\}$ with cumulative risk factor process 
$Z_i(t)=\sum_{j\le M_i(t)}\theta_iG_{i,j}$. Again, 
$x_i$ may or may not enter parameters of $M_i$, $\theta_i$, 
or the distribution of $G_{i,j}$. 
For a first illustration, assume that the Poisson process $M_i(\cdot)$ 
for individual $i$ has intensity 
$\lambda_i(s)=\lambda_0(s)\,\exp(\beta^\tr x_i)$,
as happens with standard Poisson regression modelling, 
and that both the $\theta_i$s as well as the risk multipliers 
$R_{i,j}=1-\exp(-G_{i,j})$ have the same distribution 
across individuals. Then (6.4) implies that individual $i$
has hazard rate function 
$$h_i(s)=\lambda_0(s)\,\exp(\beta^\tr x_i)
   \,\E\{1-\exp(-\theta G)\}. $$
In other words, the Cox regression structure 
has been derived from the frailty process model. 
For a second illustration with a less standard outcome, 
let the $\lambda_i(s)$ be as above, take the $\theta_i$s to be 1, 
and model the $R_{i,j}$ as arising from a Beta distribution 
with parameters $(c\mu(x_i),c-c\mu(x_i))$, say. 
This allows individuals with different covariates to
have different expected levels for their risk multipliers. 
A reasonable model emerging from this would be 
$$h_i(s)=\lambda_0(s)\,\exp(\beta^\tr x_i)\mu(x_i)
             =\lambda_0(s)\,\exp(\beta^\tr x_i)
              {\exp(\gamma^\tr x_i)\over 1+\exp(\gamma^\tr x_i)}, $$
for example, with additional $\gamma$ parameters to model
the $\mu(x_i)$. The point to note is that the $1-L_0(\theta)$ term 
always is inside $(0,1)$. A particular case with a reasonable biological 
basis is the one with a common Poisson rate but different 
impacts $R_j=1-\exp(-G_j)$ for different individuals, 
entailing a hazard rate structure of the form 
$h_i(s)=\lambda_0(s)\exp(\gamma^\tr x_i)/\{1+\exp(\gamma^\tr x_i)\}$. 

More general \LL{} frailty processes may also be considered here 
in the place of (6.1), and different specialisations lead
to different hazard regression structures. 
Such are developed and discussed in Gjessing, Aalen and Hjort (2003).
We should point out that also additive regression forms may be derived 
for the hazards under other assumptions for the \LL{} processes. 
The theme here is obviously of a general nature. It concerns 
the study of biologically plausible background process models, 
not immediately or not necessarily with the aim of analysing direct data 
from them, but rather to deduce plausible functional forms of important
statistical quantities. This theme is also visible in some
of the work reported on in S.~Richardson's chapter (this volume). 
Such lines of research do have a strong
tradition in statistics and probability theory, dating back 
more than a century, but have perhaps been underplayed in 
much of contemporary work. 

\section{Beta processes in a linear hazard model} 

The purpose of this section is to indicate how the Beta process,
introduced in Hjort (1984, 1990)
as a nonparametric Bayesian tool for modelling cumulative hazard rates 
in event history analysis, can be used also in Aalen's 
additive hazard regression model. 

Assume survival data exist in the form of triplets $(t_i,x_i,\delta_i)$ 
for $n$ individuals, where $t_i$s are life-times, possibly censored, 
the $x_i$s are covariates of dimension $p$, 
and the $\delta_i$s are indicators for non-censoring. 
In contrast to the multiplicative Cox regression model,
Aalen's linear hazard regression model takes an additive form
$h_i(s)=\alpha_0(s)+x_{i,1}\alpha_1(s)+\cdots+x_{i,p}\alpha_p(s)$, 
with a consequent expression for the cumulative hazard rate $H_i$ 
and for survival distributions 
$$S(t\midd x_i)=\exp\{-H_i(t)\}
  =\bar G_0(t)\bar G_1(t)^{x_{i,1}}\cdots\bar G_p(t)^{x_{i,p}} \eqno(7.1)$$
for an individual with covariate vector $x_i$, where $\bar G_j=1-G_j$ 
is the survival function having $\alpha_j$ as hazard rate. 
This model is typically analysed nonparametrically, 
with emphasis on Aalen plots for the cumulative risk factor functions; 
see Aalen (1989). We will use \LL{} processes for some of these components, 
and need a framework able to handle discrete cumulative hazard rates
as a function of continuous time. The canonical model formulation is 
that the cumulative hazard $H(t\midd x_i)$ for an individual with
covariate information $x_i$ should have increments obeying 
$$1-\d H(s\midd x_i)
  =(1-\d A_0(s))(1-\d A_1(s))^{x_{i,1}}\cdots(1-\d A_p(s))^{x_{i,p}} $$
for all $s$. This implies (7.1) again, 
with $\bar G_j(t)=\prod_{[0,t]}(1-\d A_j(s))$;
see e.g.~Hjort (1990) for the product integral.

One may now attempt nonparametric Bayesian modelling of the $A_j$
or $G_j$ functions inside this framework. In the general Aalen model
these increments are allowed to be both positive and negative
(as long as (7.1) behaves like a survival function),
and a possibility is to use $A_j=B_j-C_j$ where $B_j$ and $C_j$
are independent \LL{} processes with nonnegative infinitesimal 
increments bounded by 1. 
Let us here focus on the separate submodel where the $A_j$s 
are to have nonnegative increments. This is a meaningful model
when the $x_{i,j}$s represent risk levels associated with risk factors 
that a priori increase the hazard. This submodel is particularly
suited for the case where normal and healthy individuals follow
a life-time distribution governed by $A_0$, corresponding to
each $x_{i,j}=0$, and where increased $x_{i,j}$s means increased hazard. 
 
In such a situation a natural prior takes the form of 
independent Beta processes $\Beta(c_j,A_{0,j})$ for $A_j$;
the increments $\d A_j(s)$ are independent and approximately 
Beta distributed with mean $\d A_{0,j}(s)$ and variance 
$\d A_{0,j}(s)/\{1+c_j(s)\}$. One now needs to generalise the 
original main theorem about Beta processes to arrive at the  
posterior distribution of $A_0,\ldots,A_p$ given a set of 
$(y_i,x_i,\delta_i)$ data. Such a result has been derived 
in an unpublished technical report from 1997. 
Essentially, the $A_j$s still behave like Beta processes 
(with updated parameters) between observed data points 
$t_{(1)}<\cdots<t_{(n)}$, and there are jumps $\Delta A_j(t_{(i)})$ 
at the data points with a certain non-standard distribution. 
Formulae for $\hatt A_j=\E(A_j\midd\data)$ have been obtained, 
likewise for $\hatt S(t\midd x)=\Pr\{T\ge t\midd x,\data\}$
and for posterior variances and covariances. 
One may also simulate from the posterior,
to allow Bayesian inference for all parameters of interest. 
Such a programme has been carried out in Beck (2000). 

\section{Local Bayesian regression} 

In this section we study a class of Bayesian non- and semiparametric 
methods for estimating regression curves and surfaces. 
The main idea is to model the regression as locally linear,
and then place suitable local priors on the local parameters.

The nonparametric regression problem concerns data 
$Y_i=m(x_i)+\eps_i$ for $i=1,\ldots,n$ where the $\eps_i$s 
are zero-mean i.i.d.~with standard deviation $\sigma$, 
and where the $m(x)$ function is unknown. 
The favoured frequentist method is that of local polynomials,
of which special cases are the `local constant' and the `local linear'
methods; see Fan and Gijbels (1996) for an exposition. 
The local linear method minimises for given position $x$ 
the function $\sumin K_h(x_i-x)\{Y_i-(a+b(x_i-x))\}^2$ w.r.t.~$(a,b)$, 
and uses $\tilda m(x)=\tilda a=\tilda a_x$ as estimator.
Here $K_h(u)=h^{-1}K(h^{-1}u)$ is a scaled version 
of a kernel function $K(u)$. 

Bayesian nonparametric regression must involve prior modelling 
of the curve $m(x)$ and calculations related to its posterior.
The spline smoothing apparatus may be phrased in such terms.
Here we outline methods which become Bayesian generalisations
of the successful local polynomial modelling strategy. 
For illustration of the general ideas we focus here on the `local constant' 
model and method. The classical Nadaraya--Watson estimator is 
the minimiser $\tilda m(x)$ of $\sumin K_h(x_i-x)(Y_i-a)^2$, which is 
$\sumin \bar K(h^{-1}(x_i-x))Y_i/\sumin \bar K(h^{-1}(x_i-x))$, 
where we take $\bar K(u)=K(u)/K(0)$ to be a symmetric, unimodal 
kernel function, supported on $[-\half,\half]$. Now consider 
$$L_n(x,a,\sigma)=\prod_{i\in N(x)}f(y_i\midd x_i,a,\sigma)
  ^{\bar K(h^{-1}(x_i-x))}, \eqno(8.1)$$
where $Y_i\midd x_i\sim \N(a,\sigma^2)$ and the product 
is over a local neigbourhood $N(x)=[x-\half h,x+\half h]$. 
The local likelihood view is to interpret $\bar K(h^{-1}(x_i-x))$ 
as the information weight carried by data pair $(x_i,y_i)$ 
for the local $a=a_x$ parameter. Maximisation of (8.1) gives
the local likelihood estimator, which is also the local constant
estimator, and for this operation the level of $\bar K$ 
is immaterial; $\bar K$ and $K$ give the same results. 
For the local Bayesian computation we insist on using $\bar K$, 
however, with maximum value 1 corresponding to having full information 
weight for the underlying model, here, the $\N(a,\sigma^2)$ model. 
The scaled kernel smooths the information value down to zero
for data pairs outside the $x\pm\half h$ window.
Note that $L_n(x,a,\sigma)$ is the genuine likelihood 
for the model over this window when the kernel is uniform.  

The local Bayesian computation starts out with a prior 
for the local parameter, say $a=a_x$, for which we take 
the prior $\N(m_0(x),\sigma^2/w_0(x))$ with a suitable local 
precision function $w_0(x)$. This prior is then combined 
with the local likelihood $L_n(x,a,\sigma)$, which is proportional to 
$\sigma^{-s_0(x)} \exp\{-\half Q(x,a)/\sigma^2\}$.
Here $s_0(x)=\sum_{i\in N(x)}\bar K(h^{-1}(x_i-x))$, which may also
be expressed as $nhf_n(x)/K(0)$ in terms of the kernel density estimator
$f_n$ based on $K$, while 
$Q(x,a)=Q_0(x)+s_0(x)\{a-\tilda m(x)\}^2$, 
in which $Q_0(x)=\sum_{i\in N(x)}\bar K(h^{-1}(x_i-x))\{y_i-\tilda m(x)\}^2$. 
The result is 
$$m(x)\midd{\rm local\ data},\sigma\sim\N\Bigl(\hatt m(x),
  {\sigma^2\over w_0(x)+s_0(x)}\Bigr), \eqno(8.2)$$
with local Bayes estimator 
$$\hatt m(x)={w_0(x)\over w_0(x)+s_0(x)}m_0(x)
   +{s_0(x)\over w_0(x)+s_0(x)}\tilda m(x). $$  
Note that the non-informative prior case
$\sigma^2/w_0(x)=\infty$ yields the frequentist local linear estimator. 

This is `so far, so good', and suffices if one really 
can come up with a prior guess curve $m_0(x)$ and 
a strength of belief function $w_0(x)$. More realistically
these are not fully specified a priori, and a more general 
local Bayesian regression programme would comprise the following steps. 
(A) Give a prior guess function $m_0(\cdot)$ and a prior for $\sigma$. 
(B) For each $x$, use the local prior $a_x\sim \N(m_0(x),\sigma^2/w_0(x))$
for the local constant $a_x$. 
(C) Do the local Bayesian prior to posterior calculation,
employing the local likelihood. This is the calculation
carried out above, with general result 
$$\hatt m(x)=\E\,(a_x\midd{\rm local\ data})
   =\hatt m(x\midd w_0(\cdot),m_0(\cdot)). $$
(D) Use empirical Bayes methods to estimate or fine-tune $w_0(x)$, 
given $m_0(\cdot)$. 
(E) Use finally hierarchical Bayes methods, involving a background or 
first-stage prior on $m_0(\cdot)=m_0(\cdot,\xi)$, say, to arrive at 
\begin{equation}
\begin{split} 
\hatt m(x)
&= \E\bigl[\hatt m(x\midd\hatt w_0(\cdot,\xi),m_0(\cdot,\xi))
        \midd{\rm all\ data}\bigr] \cr
&= \int\hatt m(x\midd\hatt w_0(\cdot,\xi),m_0(\cdot,\xi))
        \,\d\pi(\xi\midd{\rm all\ data}). 
\nonumber 
\end{split} 
\end{equation} 
This would typically be computed via simulations of $\xi_j$s
from $\d\pi(\xi\midd{\rm all\ data})$; for each of these 
one computes the precision function $\hatt w_0(x,\xi_j)$
using empirical Bayes methods, giving via (8.2) a full curve 
$\hatt m(x\midd \hatt w_0(\cdot,\xi_j),m_0(\cdot,\xi_j))$. 
In the end one averages these curves to display the curve estimate. 

We note that the computation leading to (8.2) and the Bayes estimator,
corresponding to steps (A), (B), (C), requires only studying the situation 
at a single position $x$ at a time, so to speak. Steps (D) and (E)
really require fuller simultaneous aspects of the prior modelling
of the curve, however. 
A fuller description of the local constant 
prior used to exemplify the general scheme here is that the 
curve is constant on each of many windows of length $h$, with
a simultaneous multinormal prior for the levels at these windows. 
For the local linear version of the scheme, the prior model takes
the view that the curve is approximately linear inside each of 
many small windows, with a simultaneous multinormal prior for 
the collection of local levels and local slopes. 
Details, discussion and generalisations of the various ingredients 
at work here are in Hjort (1998). 

Observe that when the width of the local data window is large 
these methods reduce to familiar fully parametric Bayesian methods,
whereas they are essentially nonparametric when the width is small.
The apparatus also encompasses the possibility of using non-informative 
reference priors for the local parameters, in which case estimators 
coincide with the by now classical local polynomial frequentist methods. 

\section{Random shapes with smoothed Dirichlets} 

Consider the class of closed curves in the plane 
which can be represented as 
$R(s)\,(\cos(2\pi s),\allowbreak \sin(2\pi s))$ for $0\le s\le 1$, 
with $R(s)$ being some smooth positive function with $R(1)=R(0)$. 
Various stochastic process models for the radius function $R(s)$
give rise to different random shape models. 
Kent, Dryden and Anderson (2000) in effect use such an approach, 
based on a circularly symmetric Gau{\ss}ian process for $R(s)$,
following up earlier work by Grenander and Miller (1994). 
This works, but is moderately unsatisfactory in that the paths 
of such a process can come below zero. This also leads 
to some interpretational and statistical problems with 
the Gau\ss ian likelihood approach used in these papers. 

A different approach which avoids some of these difficulties
is to use smoothed Gamma and Dirichlet processes for the random radius
function. Let $g_0$ be a smooth density on $[0,1]$, 
periodic in the sense that $g_0(0)=g_0(1)$, with cumulative 
distribution $G_0$. Consider a Gamma process $G$ with parameters 
$(bG_0,b)$ on $[0,1]$; in particular, $\E\,\d G(s)=g_0(s)\,\d s$ 
while $\Var\,\d G(s)=g_0(s)\,\d s/b$. 
A fruitful model is the one smoothing locally over these 
Gamma process increments. Consider therefore $R(s)=\int K_a(s-u)\,\d G(u)$ 
where $K(u)$ is a kernel probability density,
taken to be continuous and symmetric on its support $[-\half,\half]$,
and where $K_a(u)=a^{-1}K(a^{-1}u)$ for a bandwidth
parameter $a$. The radius integral is taken to be modulo the circle around 
which it lives, that is, clockwise modulo its parameter interval $[0,1]$. 
For pure shape analysis it makes sense to strip away 
any information about the size of the objects studied.
Such size normalisation can be achieved in several ways,
but the most natural strategy here is to normalise by average 
radius length, or, in other words, to condition on the event 
$\int_0^1R(s)\,\d s=G(1)=1$. Let therefore $\bar G(\cdot)=G(\cdot)/G(1)$,
which is a Dirichlet $(b,G_0)$, and $\bar R(s)=\int K_a(s-u)\,\d \bar G(u)$. 
This smoothed Dirichlet process is guaranteed to have total 
average radius length 1. The distribution of a set of random radii 
is quite complicated, but in principle determined via
the Hilbert transform results mentioned in Section 3. 

Various models of interest emerge via the use of different 
$g_0$ functions, perhaps para\-metrised to reflect wished-for 
aspects of the shapes. Kent, Dryden and Anderson (2000) focus 
on `featureless' objects. This translates into requirements 
of circular symmetry and independence of starting point and 
leads to choosing the uniform density for $g_0$. 
Thus we have a two-parameter model for a random shape, 
centred at the unit circle. Parameter $b$ has to do 
with the concentration of the Gamma increments 
around their expected values while parameter $a$ reflects the degree 
of smoothing of the independent Gamma increments. 
Parameters $a$ and $b$ need to be estimated from one or more observed shapes. 
Useful properties include the formula 
$\pi\{1+(b+1)^{-1}\int(K_a-1)^2\,\d u\}$ for the mean of the 
random area of the curve and formulae for 
${\rm cov}\{\bar R(s),\bar R(s+h)\}$. 
In ongoing work I have used empirical covariance functions and a certain 
maximum simulated likelihood strategy to determine 
parameter estimates. 

\section{Nonparametric envelopes around parametric models} 

Some nonparametric Bayesian constructions can be
viewed as providing `nonparametric envelopes' around 
traditional parametric models. In this light traditional
parametric inference is the limiting case of zero envelope width.
The nonparametric Bayesian solutions may hence be seen 
as robustifications of such procedures, allowing for 
some amount of modelling error. 

\subsection{A semiparametric Bayes model.} 

Consider a regression situation with $Y_i=x_i^\tr\beta+\sigma\eps_i$
for $i=1,\ldots,n$, where the $\eps_i$s come from a distribution $G$.
Study the prior where $(\beta,\sigma)$ has some prior density $\pi$ 
and $G$ independently comes from a $\Dir(b,G_0)$, where $G_0$ is
the standard normal. A large value of $b$ corresponds to $G$ being 
very close to $G_0$ and hence to the traditional parametric setup.
Seeing the data and knowing the parameters amounts to knowing 
the $\eps_i$s, so $G$ given data and $(\beta,\sigma)$
is an updated Dirichlet with total parameter 
$bG_0+\sumin\delta(\sigma^{-1}(y_i-x_i^\tr\beta))$. 
One may show that the posterior of the parameters is 
$\pi(\beta,\sigma\midd\data)=c\,\pi(\beta,\sigma)L_n(\beta,\sigma)$,
where $L_n$ is the likelihood under the null model $G=G_0$,
that is, the posterior is the same as it would be under the null model. 
This assumes that the $y_i$s are distinct. 
Inference for quantities that depend also on $G$ are affected
by the nonparametric part of the prior, however. 
In particular, 
$$\hatt G(t)=\E\{G(t)\midd\data\}=w_nG_0(t)
   +(1-w_n)n^{-1}\sumin
   \Pr\{\sigma^{-1}(y_i-x_i^\tr\beta)\le t\midd\data\}, $$
where $w_n=b/(b+n)$. This may be seen to be the integral of 
a smooth function $\hatt g(t)$, which is a convex combination
of the normal prior density $g_0(t)$ and a kernel-type density 
estimator $g_n(t)$ with a variable bandwidth approximately 
proportional to $n^{-1/2}$. Interestingly, a very similar story
emerges with the more general process studied in Section 2.3.
Essentially, the formula for $\hatt G$ holds but with
a value of $w_n$ being determined by the distribution $H$ of the $B_j$s.

\subsection{Model fitting with control sets} 

The extra randomness around the normal model introduced by the 
Dirichlet in the semiparametric setup above did not influence 
on the posterior distribution of $(\beta,\sigma)$. 
Suppose now that $G$ is taken to be a $\Dir(b,G_0)$,
but conditioned to have $G(B_j)=z_j$ for each of chosen sets 
$B_1,\ldots,B_k$ partitioning the sample space.
With such a pinning down of the Dirichlet the posterior becomes 
proportional to $\pi(\theta)L_n(\theta)M_n(\theta)$,
where $M_n(\theta)=\prod_{j=1}^k(bz_j)^{N_j(\theta)}/(bz_j)^{[N_j(\theta)]}$,
writing $\theta$ for $(\beta,\sigma)$. Here $N_j(\theta)$ is 
the number of $r_i(\theta)=\sigma^{-1}(y_i-x_i^\tr\beta)$ in $B_j$,
and $x^{[m]}=x(x+1)\cdots(x+m-1)$. This leads to non-standard
asymptotics for Bayes estimators, as $M_n$ is of the same
stochastic order as $L_n$; see Hjort (1986). The Bayes estimator
balances two aims of equal importance, to be close to the maximum 
of the likelihood, and to come close to having a fraction of $z_j$ 
residuals $r_i(\theta)$ in set $B_j$ for $j=1,\ldots,k$. 
This apparatus may be used when one of the intentions of fitting
a model is to predict frequencies for certain sets, and can
be tailor-made to model-robust quantile regression, for example. 

\subsection{Randomness around a parametric survival data model} 

Assume survival data of familiar type $(t_i,\delta_i)$ are 
available, where $\delta_i$ is indicator for non-censoring,
and let $\alpha_\theta(s)$ describe some parametric model 
for the hazard rate function. To create model uncertainty around it,
let $A$ be a $\Beta(c,A_0)$ process centred at the unit rate model;
its cumulative hazard rate mean is $A_0(t)=t$ and its 
variance is $t/\{1+c(t)\}$. Now postulate that 
$1-\d A_\theta(s)=\{1-\d A(s)\}^{\alpha_\theta(s)}$ for positive $s$
and give $\theta$ a prior $\pi$. For large $c$ this becomes
ordinary parametric inference for the $\alpha_\theta$ model,
while for moderate or small $c$ we have a semiparametric 
Bayesian model around the given parametric one. The survival 
function for given $\theta$ and $A$ is 
$S_\theta(t)=\prod_{[0,t]}\{1-\d A(s)\}^{\alpha_\theta(s)}$.
Here one may show that the posterior density of $\theta$ becomes
proportional to $\pi(\theta)L_n^*(\theta)$, where 
\begin{equation}
\begin{split} 
L_n^*(\theta)
&= \prod_{i\colon\delta_i=1}
   \Bigl[\psi\bigl(c(t_i)+\alpha_\theta(t_i)Y(t_i)\bigr)
      -\psi\bigl(c(t_i)+\alpha_\theta(t_i)(Y(t_i)-1)\bigr)\Bigr] \cr
&  \qquad\times
   \exp\Bigl[-\int_0^\infty\Bigl\{\psi\bigl(c(s)+\alpha_\theta(s)Y(s)\bigr)
   -\psi\bigl(c(s)\bigr)\Bigr\}\,c(s)\,\d A_0(s)\Bigr], 
\nonumber 
\end{split}
\end{equation} 
in terms of $Y(s)=\sumin I\{t_i\ge s\}$, and where 
$\psi$ is the derivative of the logarithmic gamma function. 
When $c\arr\infty$ this can be shown to become the familiar 
likelihood $L_n(\theta)$.
For moderate and smaller values of $c$ this leads to model-robust
Bayesian parametric inference. 

\section{Concluding remarks} 

This chapter has hopefully helped illustrate that the field of 
Bayesian nonparametrics is rich in challenges and possibilities.
That its reach is expanding is witnessed for example 
by the breadth of discussion contributions to 
Walker, Damien, Laud and Smith (1999). 
Also, several other chapters in this volume touch 
in various ways aspects of nonparametric Bayes. 
That its future looks bright is also helped by computational 
advances over the last decade. 

Several of the nonparametric Bayesian stories told in brevity here 
have interesting extensions to more general settings. In particular,
many of the models, methods and results surveyed above for 
the i.i.d.~situation can be generalised to situations with 
covariate information. For example, forthcoming work with
Petrone uses the quantile-Dirichlet process to develop Bayesian 
inference methods for quantile regression. Such methods 
have also been developed by Kottas and Gelfand (2001). 
The Bayesian modelling of local parameters used in Section 8
is also clearly of a general nature, and can be used 
for example to develop Bayesian Poisson regression methods.

One may also point to further challenges for the field.
A theme of interest is to build models that take 
prior notions of shape into account, like unimodality
in density estimation; see Hansen and Lauritzen (2002)
for an interesting construction. 
Another line of research is that exemplified in Section 10,
enveloping frequently used parametric models in bigger models 
via Bayesian modelling of uncertainty. This may lead to 
model-robust inference methods with clear interpretations.  
One example could be to build a time series model where 
the autocorrelation function is a nonparametrically modelled 
function centred at say the parametric ${\rm AR}(p)$ structure,
with an extra parameter to dictate the degree of closeness
to this centre function. Similar attempts could be geared towards
modelling the covariance function in geostatistical models. 
Yet further challenges include constructing and polishing 
Bayesian extensions of generalised linear models via modelling 
of the link functions, as exemplified e.g.~in Gelfand and Mallick (1995)
who used mixtures of betas to model the covariate link function
for proportional hazards. 

\subsection*{Acknowledgements} 
I have benefitted on many levels from my involvement with
the HSSS programme, also regarding stimulus for work 
reported on in this chapter. 
I have been privileged to work on these themes with 
Beno{\^\i}t Beck, 
Arnoldo Frigessi, 
H{\aa}kon Gjessing, 
Andrea Ongaro, 
Sonia Petrone, 
Jean-Marie Rolin, 
Stephen Walker and
Odd Aalen.
Thanks are also due to my fellow editors and to Natal'ya Tsilevich
for particularly constructive comments 
on an earlier version of this chapter.

\section*{References}

\begin{list}{}{\setlength{\itemindent}{-\leftmargin}}

\font\smallrm=cmr8
\font\smallsl=cmsl8
\font\smallbf=cmbx8

\item Barron, A., Schervish, M.J.~and Wasserman, L. (1999). 
The consistency of distributions in nonparametric problems.
{\sl Annals of Statistics} {\bf 27}, 536--561.

\item Beck, B. (2000).
{\sl Nonparametric Bayesian Analysis for Special Patterns
of Incompleteness.} 
Ph.D.~thesis, Department of Statistics, 
Universit\'e Catho\-lique de Louvain.

\item Billingsley, P. (1995).
{\sl Probability and Measure} (3rd ed.).
Wiley, New York. 

\item Cheng, C. (1995). The Bernstein polynomial estimator
of a smooth quantile function.
{\sl Statistics and Probability Letters} {\bf 24}, 321--330.

\item Cifarelli, D.M.~and Regazzini, E. (1990).
Distribution functions of means of a Dirichlet process.
{\sl Annals of Statistics} {\bf 18}, 429--442;
corrigendum, ibid.\allowbreak~(1994) {\bf 22}, 1633--1634.

\item Dey, D., M\"uller, P.~and Sinha, D. (1998).
{\sl Practical Nonparametric and Semiparametric Bayesian Statistics.}
Springer-Verlag, New York.

\item Diaconis, P.~and Kemperman, J. (1996). 
Some new tools for Dirichlet priors.
In {\sl Bayesian Statistics 5}
(eds.~J.M.~Bernardo, J.O.~Berger, A.P.~Dawid and A.F.M.~Smith), 97--106. 
Oxford University Press, Oxford.

\item Doksum, K.A. (1974).
Tailfree and neutral random probabilities and their posterior distributions.
{\sl Annals of Probability} {\bf 2}, 183--201.

\item Efron, B. (2002).
Robbins, empirical Bayes, and microarrays.
{\sl Annals of Statistics}, to appear. 

\item Escobar, M.D.~and West, M. (1995).
Bayesian density estimation and inference using mixtures.
{\sl Journal of the American Statistical Association} 
{\bf 90}, 577--588.

\item Fan, J.~and Gijbels, I. (1996).
{\sl Local Polynomial Modelling and its Applications.}
Chapman and Hall, London.

\item Ferguson, T.S. (1973).
A Bayesian analysis of some nonparametric problems.
{\sl Annals of Statistics} {\bf 1}, 209--230.

\item Ferguson, T.S. (1974).
Prior distributions on spaces of probability measures.
{\sl Annals of Statistics} {\bf 2}, 615--629.

\item Gelfand, A.E.~and Mallick, B.K. (1995).
Bayesian analysis of proportional hazards models built 
from monotone functions. 
{\sl Biometrics} {\bf 51}, 843--852. 

\item Ghosal, S., Ghosh, J.K.~and Ramamoorthi, R.V. (1999). 
Posterior consistency of Dirichlet mixtures in density estimation.
{\sl Annals of Statistics} {\bf 27}, 143--158.

\item Ghosal, S.~and van der Vaart, A. (2000).
Rates of convergence for Bayes and maximum likelihood estimation
for mixtures of normal densities. 
Research Report, Vrije Universiteit Amsterdam.

\item Gjessing, H.K., Aalen, O.O.~and Hjort, N.L. (2003). 
Frailty models based on L\'evy processes.
{\sl Advances in Applied Probability} {\bf 35}, 532--550. 

\item Green, P.J.~and Richardson, S. (2001).
Modelling heterogeneity with and without the Dirichlet process.
{\sl Scandinavian Journal of Statistics} {\bf 28}, 355--375.

\item Grenander, U.~and Miller, M.I. (1994).
Representations of knowledge in complex systems 
(with discussion). 
{\sl Journal of the Royal Statistical Society} 
{\bf B 56}, 549--603.

\item Guglielmi, A.~and Tweedie, R.L. (2000).
MCMC estimation of the law of the mean of a Dirichlet process.
Technical report TR 00.15, CNR--IAMI, Milano.

\item Hansen, M.B.~and Lauritzen, S.L. (2002). 
Non-parametric Bayes inference for concave distribution functions. 
{\sl Statistica Neerlandica} {\bf 56}, 110--127. 

\item Hjort, N.L. (1984).
Contribution to the discussion of Andersen and Borgan's 
`Counting process models for life history data: a review'. 
{\sl Scandinavian Journal of Statistics} {\bf 12}, 141--150.

\item Hjort, N.L. (1986). 
Contribution to the discussion of Diaconis and Freedman's 
`On the consistency of Bayes estimates'. 
{\sl Annals of Statistics} {\bf 14}, 49--55.

\item Hjort, N.L. (1990).
Nonparametric Bayes estimators based on Beta processes 
in models for life history data. 
{\sl Annals of Statistics} {\bf 18}, 1259--1294.

\item Hjort, N.L. (1996). 
Bayesian approaches to semiparametric density estimation
(with discussion contributions).
In {\sl Bayesian Statistics 5}, 
proceedings of the Fifth International Val\`encia Meeting 
on Bayesian Statistics (eds.~J.~Berger, J.~Bernardo, 
A.P.~Dawid, A.F.M.~Smith), 223--253.

\item Hjort, N.L. (1998). 
Local Bayesian regression. 
Statistical Research Report, Department of Mathematics,
University of Oslo. 

\item Hjort, N.L. (2000).
Bayesian analysis for a generalised Dirichlet process prior.
Submitted for publication.

\item Hjort, N.L.~and Ongaro, A. (2002).
On the distribution of random Dirichlet means.
Statistical Research Report, University of Oslo.

\item Hjort, N.L.~and Walker, S.G. (2001).
Nonparametric Bayesian quantile inference.
Statistical Research Report, University of Oslo.

\item Ishwaran, H.~and Zarepour, M. (2000).
Markov chain Monte Carlo in approximate Dirichlet and beta two-parameter
process hierarchical models.
{\sl Biometrika} {\bf 87}, 353--369. 

\item Kent, J.K., Dryden, I.~and Anderson, C.R. (2000). 
Using circulant symmetry to model featureless objects.
{\sl Biometrika} {\bf 87}, 527--544.

\item Kerov, A.~and Tsilevich, N. (1998).
The Markov--Krein correspondence in several dimensions.
PDMI preprint 1. 

\item Kingman, J.F.C. (1975).
Random discrete distributions.
{\sl Journal of the Royal Statistical Society} {\bf B 37}, 1--22.

\item Kottas, A.~and Gelfand, A. (2001).
Bayesian semiparametric median regression modeling.
{\sl Journal of the American Statistical Association} {\bf 96}, 1458--1468. 

\item Kraft, C.H. (1964). 
A class of distribution function processes which have derivatives. 
{\sl Journal of Applied Probability} {\bf 1}, 385--388.

\item Lavine, M. (1992).
Some aspects of Polya tree distributions for statistical modeling.
{\sl Annals of Statistics} {\bf 20}, 1222--1235.

\item Lavine, M. (1994). 
More aspects of Polya tree distributions for statistical modeling. 
{\sl Annals of Statistics} {\bf 22}, 1161--1176.

\item Lo, A.Y. (1984). 
On a class of Bayesian nonparametric estimates: I, density estimates.
{\sl Annals of Statistics} {\bf 12}, 351--357. 

\item Markov, A.A. (1896). 
Nouvelles applications des fractions continues.
{\sl Mathematische Annalen} {\bf 47}, 579--597. 

\item Regazzini, E., Guglielmi, A.~and di Nunno, G. (2000).
Theory and numerical analysis for exact distributions
of functionals of a Dirichlet process.
Research report, Universit\`a di Pavia. 

\item Rubin, D.B. (1981).
The Bayesian bootstrap.
{\sl Annals of Statistics} {\bf 9}, 130--134. 

\item Schwartz, L. (1965). 
On Bayes procedures.
{\sl Zeitschrift f\"ur Wahrscheinlichkeitstheorie und Verwandte Gebiete} 
{\bf 4}, 10--26. 

\item Sethuraman, J. (1994).
A constructive definition of Dirichlet priors.
{\sl Statistica Sinica} {\bf 4}, 639--650. 

\item Sethuraman, J.~and Tiwari, R. (1982).
Convergence of Dirichlet measures and the interpretation of their
parameter. In {\sl Proceedings of the Third
Purdue Symposium on Statistical Decision Theory and Related Topics}
(eds.~S.S.\allowbreak~Gupta and J.~Berger), 305--315. Academic Press, New York.

\item Shen, X.~and Wasserman, L. (2000).
Rates of convergence of posterior distributions.
{\sl Annals of Statistics}, to appear. 

\item Tsilevich, N.V., Vershik, A.~and Yor, M. (2000).
Distinguished properties of the gamma process, and related topics.
Pr\'epublication du Laboratoire de Probabilit\'es 
et Mod\`eles Al\'eatoires, no.~575.

\item Walker, S.G. (2000).
A note on consistency from a Bayesian perspective. 
Manu\-script, Department of Mathematical Sciences, 
University of Bath.

\item Walker, S.G., Damien, P., Laud, P.W.~and Smith, A.F.M. (1999). 
Bayesian nonparametric inference for random distributions and related
functions (with discussion). 
{\sl Journal of the Royal Statistical Society} {\bf B 61}, 485--528. 

\item Walker, S.G.~and Hjort, N.L. (2001).
On Bayesian consistency.
{\sl Journal of the Royal Statistical Society} {\bf B 63}, 811--821.

\item Wasserman, L. (1998). 
Asymptotic properties of nonparametric Bayesian procedures. 
In {\sl Practical Nonparametric and Semiparametric Bayesian
Statistics} (eds.~D.~Dey, P.~M\"uller and D.~Sinha), 293--304. 
{\sl Lecture Notes in Statistics}, Springer.

\item Aalen, O.O. (1989). 
A linear regression model for the analysis of life times.
{\sl Statistics in Medicine} {\bf 8}, 907--925.

\item Aalen, O.O.~and Hjort, N.L. (2002).
Frailty models that yield proportional hazards.
{\sl Statistics and Probability Letters} {\bf 58}, 335--342. 

\end{list}
\end{document}